\documentclass[aps,prl,reprint,showpacs,showkeys,superscriptaddress]{revtex4-1}
\usepackage{amsmath,bm,graphicx,hyperref}
\usepackage{amssymb}
\usepackage{mathrsfs}
\usepackage{times}

\usepackage{color}

\newcommand{\beq}{\begin{equation}}
\newcommand{\eeq}{\end{equation}}

\begin{document}

\title{Generalized Efimov effect in one dimension}

\author{Sergej Moroz}
\email{sergej.moroz@colorado.edu}
\affiliation{Department of Physics, University of Colorado, Boulder, Colorado 80309, USA}
\affiliation{Center for Theory of Quantum Matter, University of Colorado, Boulder, Colorado 80309, USA}
\author{Jos\'e P. D'Incao}
\email{jpdincao@jila.colorado.edu}
\affiliation{Department of Physics, University of Colorado, Boulder, Colorado 80309, USA}
\affiliation{JILA, University of Colorado and NIST, Boulder, Colorado 80309-0440, USA}
\author{Dmitry S. Petrov}
\email{dmitry.petrov@u-psud.fr}
\affiliation{LPTMS, CNRS, Univ. Paris-Sud, Universit\'e Paris-Saclay, 91405 Orsay, France}

\begin{abstract}
We study a one-dimensional quantum problem of two particles interacting with a third one via a scale-invariant subcritically attractive inverse square potential, which can be realized, for example, in a mixture of dipoles and charges confined to one dimension. We find that above a critical mass ratio, this version of the Calogero problem exhibits the generalized Efimov effect, the emergence of discrete scale invariance manifested by a geometric series of three-body bound states with an accumulation point at zero energy.\end{abstract}
\maketitle


In quantum mechanics three identical bosons in three dimensions interacting resonantly via a short-range two-body potential have an infinite tower of bound states, whose energy spectrum forms a geometric series near the accumulation point at zero energy. This was discovered theoretically by Vitaly Efimov in 1970 \cite{Efimov1970} and is known today as the Efimov effect. This effect is a beautiful example of \emph{few-body universality} since it is independent of the detailed form of the interaction potential provided it is tuned to the resonance  (i.e., whenever a zero-energy $s$-wave two-body bound state if formed). The Efimov effect has been extended to systems of distinguishable particles \cite{Efimov1973,Nielsen2001,Braaten2006,Petrov2012,Wang2013}, liberated from three dimensions \cite{Nishida2011} and found in other systems \cite{Nishida2012,Nishida2013}. During the last decade a number of experiments \cite{Kraemer2006a, Ferlaino2010,Huang2014b,Pires2014,Tung2014b} with cold atoms near Feshbach resonances \cite{Chin2010} verified various universal aspects related to Efimov physics--- the Efimov $^4\text{He}$ trimer has also been recently observed in \cite{Kunitski2015}--- and demonstrated the experimental capability to explore fundamental aspects of few-body systems in exotic regimes.

From a more general perspective, the most startling feature of the Efimov effect is \emph{discrete scale invariance} of the three-body problem, manifested in both bound and scattering three-body observables, that originates from \emph{continuous scale invariance} of the two-body interaction. 
It thus appears natural to us to generalize the Efimov effect to systems whose two-body interaction is not necessarily short-range and define it as \emph{the emergence of discrete scaling symmetry in a three-body problem if the particles attract each other via a two-body scale invariant potential} \footnote{If three-body discrete scale invariance appears only close to the threshold at zero energy, a short distance structure of the two-body potential does not have to be scale invariant. For example, any short-range potential tuned to a resonance is low-energy scale invariant.}.

Motivated by this broader perspective on the Efimov effect, we study a three-body problem with a two-body \emph{long-range} attractive potential of the form
\beq \label{invsq}
V(r)=-\frac{\alpha}{2\mu r^2}
\eeq
with $\mu$ being the reduced mass and $\alpha$ the dimensionless coupling constant. The potential (\ref{invsq}) is scale invariant and, at zero energy or for sufficiently small $r$, where the energy term can be neglected, in one dimension the two independent solutions of the two-body Schr\"odinger equation are the powers $r^{1/2\pm \sqrt{1/4-\alpha}}$. However, the (inevitable) breakdown of the $1/r^2$ law at small distances introduces a length scale $b$, made explicit by writing the linear combination of the two asymptotic solutions in the form 
\beq\label{eq:linearcomb}
\psi\sim \Big(\frac r b\Big) ^{1/2+\sqrt{1/4-\alpha}}-\Big(\frac r b\Big) ^{1/2-\sqrt{1/4-\alpha}}.
\eeq

For the further discussion it is crucial whether $\alpha$ is larger or smaller than $1/4$ \cite{Landau1958, Case1950, Braaten2006}. The case $\alpha>1/4$ corresponds to the fall of a particle to the center and the discrete scaling is manifest already in the two-body problem. Here, the exponents $1/2\pm\sqrt{1/4-\alpha}$ are complex conjugate, the two terms in Eq.~(\ref{eq:linearcomb}) should be treated on equal footing, and $b$ becomes an essential parameter, which can never be neglected. By contrast, the case $\alpha<1/4$ has two scale-invariant limits $b=0$ and $b^{-1}=0$ where, respectively, only the plus-branch $r^{1/2+\sqrt{1/4-\alpha}}$ or only the minus-branch $r^{1/2-\sqrt{1/4-\alpha}}$ survives in Eq.~(\ref{eq:linearcomb}). In practice, these two limits require, respectively, $|b|\ll \xi$ or $|b|\gg \xi$, where $\xi$ is a typical lengthscale in the problem such as the system size, de Broglie wave length, etc. For instance, in Eq.~(\ref{eq:linearcomb}) the minus-branch solution can be neglected if $(b/\xi)^{\sqrt{1-4\alpha}}\ll 1$.  
Thus, the plus-branch scaling is realized ``automatically'' by increasing the typical size of the system, whereas the minus-branch requires a fine tuning of the short-range part of the potential \cite{Sutherland1971a,Kaplan2009a, supmat}. Physically, this fine tuning signals the  appearance of an additional two-body bound state emerging from the zero-energy threshold which can be realized using, for example, the Feshbach resonance technique \cite{Chin2010}.

As far as the three-body problem with the two-body interaction (\ref{invsq}) is concerned,  Calogero solved it in one dimension analytically for three {\it identical} particles \cite{Calogero1969} and found continuous scale invariance for all $\alpha<1/4$, which implies the absence of the Efimov effect \cite{Guevcomment}. In this paper we show that this conclusion does not hold in general for the modified Calogero problem -- two identical spinless bosons or fermions interacting with a third particle via the potential (\ref{invsq}). In addition to the quantum statistics and the choice $b=0$ or $b^{-1}=0$ the modified problem is parametrized by the two continuous dimensionless quantities: $\alpha$ and the mass ratio. Accordingly, we calculate the critical line separating the Efimov and scale-invariant regions and describe the nature of the three-body bound state spectrum in this parameter space.

The three-body Hamiltonian relevant for our problem reads
\begin{equation}\label{model}
H=-\frac {\partial_{R_1}^2+ \partial_{R_2}^2} {2M} -\frac {\partial_r^2} {2m}+V(r-R_1)+V(r-R_2),
\end{equation}
where $R_1$ and $R_2$ are the coordinates of two identical particles of mass $M$ and $r$ is the coordinate of the third particle of mass $m$. The potential $V$ is given by Eq.~(\ref{invsq}), where $\mu=mM/(M+m)$ and $\alpha$ denotes the interspecies dimensionless coupling.

A convenient way to solve this problem is obtained using hyperspherical coordinates. First, we introduce the center-of-mass and mass-scaled Jacobi coordinates 
$R_{CM}=\big[m r+M(R_1+R_2)\big]/(2M+m)$, 
$x=\sqrt{\tilde \mu /2}(2r-R_1-R_2)$,
$y=\sqrt{2\mu_M}(R_2-R_1)$,
where $\tilde \mu=2mM/(m+2M)$ and $\mu_M=M/2$. It is then convenient to define polar (hyperspherical) coordinates
$
x=\mathcal{R} \cos\theta$, $y=\mathcal{R} \sin\theta
$
with the mass-scaled hyperradius
$
\mathcal{R}=\sqrt{2\sum_i m_i (r_i-R_{CM})^2}.
$
The interparticle distances in the new coordinates become
$r-R_1=\mathcal{R} \sin(\Delta+\theta)/\sqrt{2\mu}$, 
$r-R_2=\mathcal{R} \sin(\Delta-\theta)/\sqrt{2\mu}$ and
$R_2-R_1=\mathcal{R} \sin\theta/\sqrt{M}$,
where $\Delta=\arctan \sqrt{1+2M/m}$. Accordingly, after separating the center-of-mass motion, the relative part of the Hamiltonian \eqref{model} is written as a two-dimensional radial problem
\beq \label{rad}
H=-\partial_\mathcal{R}^2-\frac{1}{\mathcal{R}}\partial_\mathcal{R}+\frac{1}{\mathcal{R}^2} \mathcal{M}^2_{\theta}
\eeq
with the hyperangular Sch\"odinger operator
\beq \label{Mtheta2}
\mathcal{M}^2_{\theta}=-\partial_\theta^2-\frac{\alpha}{ \sin^2(\Delta+\theta)}-\frac{\alpha}{\sin^2(\Delta-\theta)}.
\eeq
Two-body scale invariance leads to \emph{separability} of the three-body problem in hyperspherical coordinates.
The relative part of the three-body wave function $\Psi(\mathcal{R},\theta)$ can thus be written in the factorized form
$
\Psi(\mathcal{R},\theta)=\Phi(\mathcal{R})\psi(\theta)
$
and the problem separates into two tasks. First, one finds $\psi$ by diagonalizing the operator $\mathcal{M}^2_{\theta}$
\beq \label{polSch}
\mathcal{M}^2_{\theta} \psi= -s^2 \psi.
\eeq 
Then, $\Phi(\mathcal{R})$ is the solution corresponding to the Hamiltonian (\ref{rad}) with $\mathcal{M}^2_{\theta}$ substituted by $-s^2$. This second task is trivially solved in terms of the Bessel functions $J_{\pm is}$, and the onset of the generalized Efimov effect coincides with the point $s^2=0$: for positive $s^2$ the system is Efimovian and for negative $s^2$ it is scale invariant. Thus, the problem of determining the critical mass ratio is equivalent to solving the hyperangular problem (\ref{polSch}) and identifying the zero crossing of $s^2$ as a function of $\Delta$. We will now discuss this procedure. 

\begin{figure}[t]
 \includegraphics[width=0.7\columnwidth,clip]{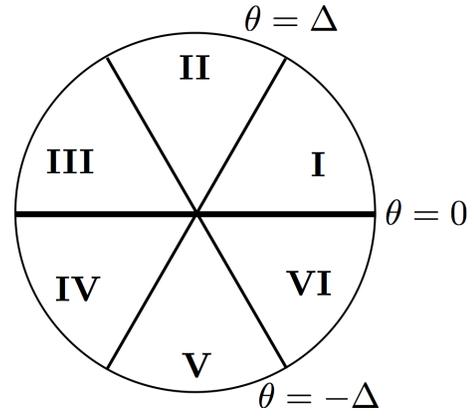}
 \caption{ Hyperangular domain partitioning.  \label{figpart}}
\end{figure}

\begin{figure}[t]
 \includegraphics[width=1.0\columnwidth,clip]{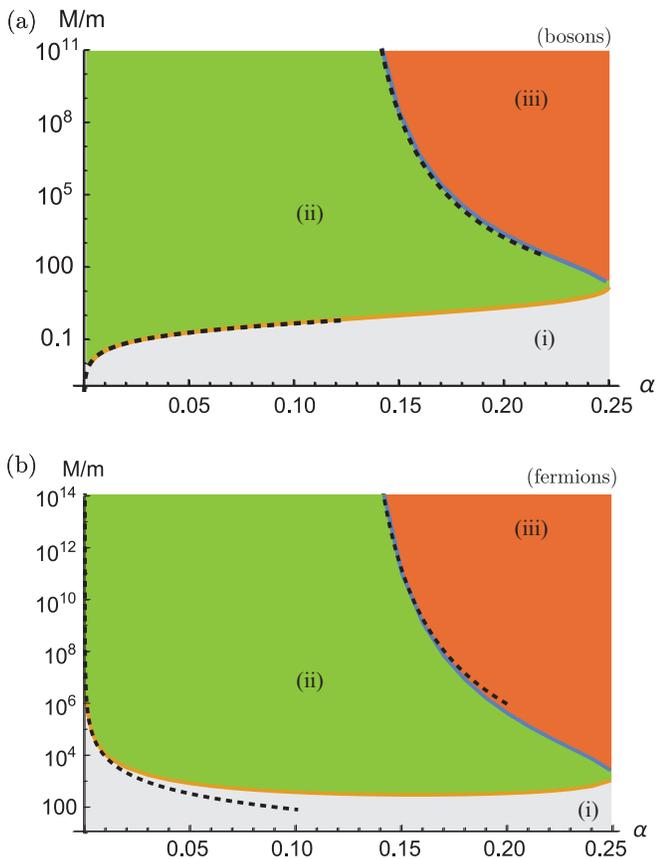}
 \caption{(Color online) (a) Critical mass ratio as a function of $\alpha$ for bosons: upper blue (lower orange) line is the Numerov numerics for the plus (minus) branch, dashed lines are analytic asymptotes near $\alpha=0$ and $\alpha=1/8$ (see text). The shaded regions (i), (ii), and (iii), denote the regimes in which the Efimov effect does not occur, occurs for the minus-branch solution only, and occurs for both minus- and plus-branch solution. (b) same as (a) but for fermions. \label{bosonsfermions}}.
\end{figure}

The coincidence angles $\theta=0, \pi$ ($M-M$ coincidence) and $\theta=\pm \Delta, \pi \pm \Delta$ ($M-m$ coincidences) partition the hyperangular circle into six regions (see Fig. \ref{figpart}).
Since two particles of mass $M$ are identical, the wave function satisfies $\psi(\theta)=\psi(-\theta)$ or $\psi(\theta)=-\psi(-\theta)$, respectively, for bosons or fermions. In addition, the hyperspherical Hamiltonian is symmetric under $\theta\to \pi-\theta$ and the wave function $\psi$ is either even or odd under this transformation.    It is thus sufficient to solve the angular problem only in the domain $\theta\in (0, \pi/2)$. Moreover, we will assume that the distinguishable particles are \emph{impenetrable}. 
Physically, this is realized by regularizing the inverse square potential \eqref{invsq} with a short-range potential that has a strong repulsive core. Due to the interspecies impenetrability, sectors I and II in Fig.~\ref{figpart} decouple and can be addressed separately. In sector I the hyperangular wave function, $\psi(\theta)$, should satisfy the following boundary conditions for $\theta= 0$ \cite{Girardeau} 
\beq \label{qstat}
 \begin{split} & \psi=0 \quad \text{fermions}, \\ & \psi' = 0 \quad \text{bosons} \end{split}
\eeq
and for $\theta\to \Delta^-$
\beq \label{branch}
 \begin{split} & \psi\sim (\Delta-\theta)^{1/2+\sqrt{1/4-\alpha}} \quad \text{plus-branch}, \\ & \psi\sim (\Delta-\theta)^{1/2-\sqrt{1/4-\alpha}} \quad \text{minus-branch}. \end{split}
\eeq
The critical mass ratio is determined by solving Eq. \eqref{polSch} in sector I and is plotted in Fig.~\ref{bosonsfermions}~(a) and (b) for bosons and fermions, respectively. We found that $s^2$ is an increasing function of the mass ratio $M/m$ for any choice of $\alpha$ and boundary conditions. In addition, we find no zero-energy ($s=0$) solution in sector II that satisfies the proper scale-invariant boundary conditions at the interspecies coincidence point. The wave function is thus zero in sector II, i.e., the probability to find the particle of mass $m$ in between the two identical particles of mass $M$ vanishes.

It should be noted that the modified Calogero problem of the type (\ref{model}) is exactly solvable and scale invariant for the plus-branch under the condition $M/m=1/(1/2+\sqrt{1/4-\alpha})$ \cite{Krivnov1982,Sen1996,Meljanac2003}. The Efimov region corresponds to higher values of $M/m$ and, since the problem is not solvable, we solve it numerically. We use the Numerov method \cite{Giannozzi2014} on a logarithmic grid (see Ref. \cite{supmat}). Nevertheless, we also find approximate analytic solutions for this problem in limiting cases discussed below.

For the plus-branch the critical mass ratio diverges at $\alpha=1/8$. In fact, for $\alpha>1/8$ both branches give rise to the Efimov effect for sufficiently large $M/m      $\footnote{This is also true in a three-dimensional version of the problem since in that case the critical value of $\alpha$, where the two-body problem becomes Efimovian, is also $1/4$. In two dimensions the critical value is exactly zero, so no three-body Efimov effect is possible.}. Indeed, for $M/m\rightarrow \infty$ the angle $\Delta=\pi/2$ and the hyperangular potential in Eq.~(\ref{Mtheta2}) reduces to $-2\alpha/\cos^2\theta$. One can see that the hyperangular problem becomes Efimovian for $2\alpha>1/4$ independent of the quantum statistics of the heavy particles and the branch choice. This means that the spectrum of $\mathcal{M}^2_{\theta}$ is unbound from below with deep bound states localized close to $\theta=\pi/2$. As a result, a finite $\pi/2-\Delta$ is necessary to renormalize this potential and bring the ground state energy $-s^2$ to zero. Quantitatively, for the plus-branch solution in the vicinity of $\alpha=1/8$ we obtain \cite{supmat}
\beq \label{alpha18}
\frac{\pi}{2}-\Delta\approx\mathscr{N} e^{\pm \frac \pi 2 -\frac{2\pi}{\sqrt{8\alpha-1}}},
\eeq
where the upper (lower) sign corresponds to the case of bosons (fermions) and $\mathscr{N}=16 \exp[-2-2\sqrt{2}-H_{(- 3+\sqrt{2})/ 2 }]\approx 11.887$ with $H_n$ being the harmonic number.
 
For the minus-branch the spectrum is Efimovian for any $0<\alpha<1/4$ for $M/m\gg 1$ \cite{supmat}. The less stringent condition for the Efimov effect in this case can be explained by the fact that the minus-branch two-body interaction nearly binds two particles and is, in this sense, more attractive than the plus-branch interaction with the same $\alpha$. In fact, the hyperangular problem can be solved analytically close to the non-interacting point $\alpha=0$. In Ref. \cite{supmat} we show that for the bosonic case
\beq \label{alpha0b}
\Delta-\frac{\pi}{4}\approx \frac{\alpha \pi}{2}
\eeq
and for fermions
\beq \label{alpha0f}
\frac{\pi}{2}-\Delta\approx \frac{\alpha \pi}{4}.
\eeq
The asymptotes (\ref{alpha18}), (\ref{alpha0b}), and (\ref{alpha0f}) are plotted in Figs. \ref{bosonsfermions} as dashed lines.

The identified critical mass ratio is calculated using the wave function $\psi(\theta)$ without nodes inside sector I. As one increases the mass ratio,
 wave functions with increasing number of nodes 
 will give rise to additional towers of Efimov states.


Now we describe the qualitative nature of the three-body bound state spectrum.
The interaction in Eq.~(\ref{invsq}) must be regularized at short distances, see \cite{supmat}. As the short-range potential depth $D_0$ changes one can tune between a pure plus-branch ($b=0$)  and minus-branch ($b^{-1}=0$) solutions. The nature of the three-body spectrum will depend on which region in Fig. \ref{bosonsfermions} the system falls into. There are three different regimes:
\begin{itemize}
\item In the region (i), below the orange curve, there is no Efimov effect for any value of $b$.
\item In the region (ii), between the orange and blue critical curves, the spectrum behaves similar to the original Efimov problem \cite{Efimov1970}. By starting from the plus branch solution with $b=0$ and increasing the depth $D_0$, three-body bound states emerge one-by-one from the three-body continuum as one approaches $b^{-1}=0$. At the critical point $b^{-1}=0$, where a zero-energy two-body bound state pops up, an infinite tower of Efimov states is formed with the Efimov parameter $s_-$ (encoding the geometric factor $e^{2\pi/s_-}$ for the energy spectrum) which depends on both $M/m$ and $\alpha$. As one further increases the depth $D_0$, the trimers disappear one-by-one into the particle-dimer continuum.
\item In the region (iii) the spectrum resembles the one appearing in the three-dimensional Efimov problem of particles with \emph{unequal} scattering lengths \cite{Efimov1973,Dincao2009,Wang2012}. Now both the plus- and minus-branches support Efimov states characterized by the Efimov parameter $s_+$ and $s_-$, respectively, where $0<s_+<s_-$. The energy spectrum contains an infinite number of three-body bound states close to the zero-energy threshold for any value of $b$. The interpolation from the plus- to the minus-branch can be understood as follows: Near $b=0$ the energy spectrum close to the zero-energy threshold is controlled by the $s_+$ parameter. As one approaches $b^{-1}=0$ the \emph{virtual} dimer state of size of order $|b|$ is formed. As a result, the trimers with energies below (above) $\epsilon_d\sim-1/{\mu b^2}$ follow the geometric scaling with the Efimov parameter $s_-$ ($s_+$). At resonance $b^{-1}=0$ the geometric spectrum with $s_-$ scaling is obtained. When the depth $D_0$ is increased further, a two-body \emph{bound} state is formed and three-body states with energies  below (above) $\epsilon_d$ will also follow the geometric scaling with the Efimov parameter $s_-$ ($s_+$) to the point where, when away enough from $b^{-1}=0$, the energy spectrum is again completely controlled by the $s_+$ parameter. 
\end{itemize}

\begin{figure}[t]
 \includegraphics[width=1.0\columnwidth,clip]{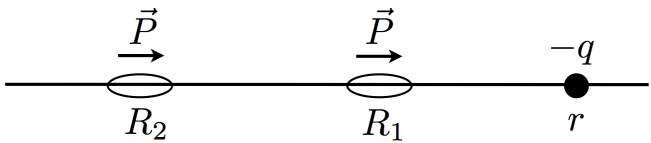}
 \caption{ Two identical dipoles and a charge confined in one dimension.  \label{dipole}}.
\end{figure}

Can the Efimov effect found in this paper be discovered experimentally? A promising candidate might be a mixture of dipoles and charges that are confined to one dimension. Indeed, the dipole-charge interaction in three dimensions is given by the scale-invariant anisotropic potential
$
V(\mathbf{r})\sim \cos \phi/r^2,
$
where $\phi$ is the angle between the direction of the dipole moment and the dipole-charge separation vector. 
Consider now a system of two identical dipoles of mass $M$ and dipole moment $P$ 
and a particle of mass $m$ and charge $-q$ confined to a one-dimensional line. Let us also regard the dipoles as dumbbells with fixed dipole moments
as illustrated in Fig. \ref{dipole}.
This three-body problem is governed by the Hamiltonian
\beq \label{modeldip}
\begin{split}
H&=-\frac 1 {2M} \left( \partial_{R_1}^2+ \partial_{R_2}^2 \right)-\frac 1 {2m} \partial_r^2 \\
&\underbrace{-K_e q P \Big[\frac 1 {(r-R_1)^2}+\frac 1 {(r-R_2)^2} \Big]}_{\text{dipole-charge}}\underbrace{- \frac {2K_e P^2} {(R_1-R_2)^3}}_{\text{dipole-dipole}},
\end{split}
\eeq
where the Coulomb constant $K_e=1/(4\pi \epsilon)$. If we neglect the dipole-dipole term, this Hamiltonian maps on Eq. \eqref{model} with
\beq \label{Peq}
\alpha=2\mu K_e P q
\eeq
and thus gives rise to the Efimov effect provided $\alpha<1/4$ and the mass ratio is above the critical value.
The presence of the dipole-dipole term introduces a length scale
$
l_{dd} \sim M K_e P^2\sim \frac P q
$
and bound states of this size. In the Efimov regime the length $l_{dd}$ provides the high-energy cutoff for the energy spectrum. This cutoff should not affect the (geometric) energy spectrum close to the zero-energy threshold. However, since our problem is one-dimensional, the dipole-dipole interaction effectively fermionizes the dipoles since their wave function is suppressed at $R_1-R_2\sim l_{dd}$. Thus, the critical mass ratio for the Efimov effect should be read off of Fig.~\ref{bosonsfermions}~(b) rather than (a) even if the dipoles are identical bosons.

As an example, consider two polar molecules interacting with an electron. From Eq. \eqref{Peq} the dipole moment of the polar molecule should satisfy $P<P_{cr}=e a_0/8\approx 0.318 \, \text{Debye}$, where $e$ is the charge of the electron and $a_0$ is the Bohr radius. Such a system has a large mass ratio, and, therefore, provided it falls into the region (iii) in Fig. \ref{bosonsfermions} (b), displays Efimov states (without fine-tuning to the minus branch) which can be detected spectroscopically. For a typical mass ratio $M/m=10^5$ we find $s_+>1$ for $P>0.281\, \text{Debye}$. Moreover, Efimov states could also be observed if 
tuning of the dipole moment $P$ is possible. In that case, near an electron-dipole
resonance, dipolar losses should be enhanced
every time a new Efimov state is formed.

A three-dimensional version of the problem may have better chances to be realized experimentally and should also have interesting quantum-chemistry implications. In that case $P_{cr}\approx1.63 \, \text{Debye}$ \cite{Connoll2007, Camblong2001, Giri2008}. 
We consider it as a promising project and leave it for future studies.

\begin{acknowledgments}
We thank S. Endo, O. Kartavtsev, Y. Nishida,  S. Tan for useful suggestions. We are grateful to V. Efimov for discussions concerning the qualitative nature of the energy spectrum. This research was partially supported by the NSF through DMR-1001240. SM is grateful for hospitality to Institute for Nuclear Theory in Seattle, where this work was partially done.
JPD acknowledges support from the U. S. National Science Foundation.
DSP acknowledges support from the IFRAF Institute. The research leading to these results has received funding from the European Research Council under European Community's Seventh Framework Programme (FR7/2007-2013 Grant Agreement no.341197).
\end{acknowledgments}

\bibliography{library_new.bib}


\clearpage

\section{Supplemental material: Generalized Efimov effect in one dimension}
\setcounter{page}{1}
\renewcommand{\theequation}{S\arabic{equation}}
\setcounter{equation}{0}
\renewcommand{\thefigure}{S\arabic{figure}}
\setcounter{figure}{0}

\subsection{Square well regularization and resonance}
Consider two particles with the reduced mass $\mu$ interacting in one dimension via the regularized inverse square potenital
\beq
\begin{split}
V(r)&=-\frac \alpha {2\mu r^2}, \quad r>r_0, \\
V(r)&=-D_0=-\frac \gamma {2\mu r_0^2}, \quad 0<r\le r_0,
\end{split}
\eeq
where $r_0>0$, $\gamma>0$. We will also assume $V(-r)=V(r)$, which implies that a wave function should be either even or odd under the reflection $r\to -r$. 
It should be noted that the regularized potential necessarily supports a parity-even ground state with energy $\epsilon_{\text{GS}}\sim - 1/(\mu r_0^2)$. This is in agreement with a general theorem that any attractive well supports at least one bound state in one dimension. Here, however, are interested in the  zero energy solution.

For $r>r_0$ a general solution of the relative Schr\"odinger equation at zero energy is given by
\beq
\psi(r)=A_+\Big(\frac {r} {r_0}\Big)^{1/2+\sqrt{1/4-\alpha}}+A_-\Big(\frac {r} {r_0}\Big)^{1/2-\sqrt{1/4-\alpha}}.
\eeq
On the other hand, for $r\le r_0$ the solution at zero energy is
\beq
\begin{split}
\psi(r)&=C \sin \Big(\sqrt{\gamma}\frac {r} {r_0}\Big) \qquad \text{reflection odd}, \\
\psi(r)&=C \cos \Big(\sqrt{\gamma}\frac {r} {r_0}\Big) \qquad \text{reflection even}. \\
\end{split}
\eeq
By matching the logarithmic derivative at $r=r_0$ one finds for the reflection odd case
\beq\label{eq:X}
X\equiv \frac {A_+} {A_-}=\frac{\sqrt{1/4-\alpha}-1/2 + \sqrt{\gamma} \cot \sqrt{\gamma}}{\sqrt{1/4-\alpha}+1/2 - \sqrt{\gamma} \cot \sqrt{\gamma}}.
\eeq
For the reflection even case one must replace $\cot \sqrt{\gamma}\to -\tan \sqrt{\gamma}$ in Eq.~(\ref{eq:X}). For a generic value of $\gamma$, the plus-branch is dominant for $r\gg r_0$ since $X\ne 0$. Note however that one can fine tune the regularized potential such that the wave function follows the minus-branch for $r\gg r_0$. This happens for $X=0$ resulting in the resonance condition
\beq
\begin{split}
\sqrt{1/4-\alpha}-1/2+\sqrt \gamma \cot \sqrt \gamma&=0 \qquad \text{reflection odd}, \\
\sqrt{1/4-\alpha}-1/2-\sqrt \gamma \tan \sqrt \gamma&=0\qquad \text{reflection even}.
\end{split}
\eeq
By increasing the depth $D_0$ of the short-range potential, the total potential supports more and more bound states. Incidentally, the resonance condition  signals the  appearance of an additional bound state emerging from the zero-energy threshold.



\subsection{Logarithmic grid and Langer substitution}
A numerical solution of the Schr\"odinger equation
\beq \label{Schgen}
[\partial_\theta^2-v(\theta)]\psi=\kappa^2 \psi
\eeq
with a potential possessing an inverse square singularity requires a grid which is dense near the singularity allowing to resolve the behavior of the wave function close to it. This can be accomplished by the Langer substitution
\beq
\theta(x)=\exp(x), \qquad \psi(\theta)= \exp\big(\frac x 2\big) y(x)
\eeq
which transforms Eq. \eqref{Schgen} into
\beq
\partial_x^2 y(x)+g(x)y(x)=0
\eeq
with
\beq
g(x)=-\exp(2x)\Big\{ \kappa^2+v[\exp(x)] \Big\}-\frac{1}{4}.
\eeq
This equation is now solved by the Numerov method on the  equidistant  $x$-grid with a lower cutoff $x_{\text{min}}<0$.

\subsection{Analytic calculation near $\alpha=1/8$}
Slightly above $\alpha=1/8$ for the plus-branch the critical mass ratio is large ($\Delta\to \pi/2$) and the angular problem can be attacked analytically. Here we sketch this calculation. First, it is convenient to redefine the angle to be $\phi=\Delta-\theta$. The angular Schr\"odinger equation \eqref{polSch} to be solved in sector I takes the form
\beq \label{schr}
\Big[-\partial_\phi^2-\frac{\alpha}{ \sin^2 \phi}-\frac{\alpha}{\sin^2(\phi+2\epsilon)}\Big] \psi=0,
\eeq  
where $\epsilon=\pi/2-\Delta\ll 1$. 

In the regime $\phi\gg \epsilon$ this equation simplifies to
\beq \label{schrl}
\Big(-\partial_\phi^2-\frac{2\alpha}{ \sin^2 \phi}\Big) \psi=0
\eeq  
with a general solution
\beq \label{soll}
\begin{split}
\psi=&(1-\cos\phi)^{\frac 1 4}  \\
& \times\Big[ L_1  P_{- 1/2}^{ \sqrt{1-8\alpha}/2}(\cos \phi)+L_2  Q_{- 1/2}^{ \sqrt{1-8\alpha}/2}(\cos \phi)\Big],
\end{split}
\eeq
where $P_n^m(x)$ and $Q_n^m(x)$ are associated Legendre functions. The quantum statistics condition \eqref{qstat} imposed at $\phi=\Delta$ fixes the ratio $L_1/L_2$.

In the regime $\phi\ll 1$ one can expand the denominators in Eq. \eqref{schr} and it transforms to
\beq \label{schrs}
\Big[-\partial_\phi^2-\frac{\alpha}{\phi^2}-\frac{\alpha}{(\phi+2\epsilon)^2}\Big] \psi=0
\eeq  
with a general solution
\beq  \label{sols}
\begin{split}
&\psi=\epsilon \sqrt{x(2+x)}  \\
& \times\Big[ S_1  P_{(\sqrt{1-8\alpha}-1)/2}^{ \sqrt{1-4\alpha}}(1+x)+S_2  Q_{(\sqrt{1-8\alpha}-1)/2}^{ \sqrt{1-4\alpha}}(1+x)\Big],
\end{split}
\eeq
where $x=\phi/\epsilon$. The ratio $S_1/S_2$ is now determined by applying the plus-branch boundary condition $\psi\sim x^{1/2+\sqrt{1/4-\alpha}}$ at $x=0^+$ [equivalent to Eq. \eqref{branch}].

In the intermediate regime $x\gg1$ and $\phi\ll 1$ the wave functions \eqref{soll} and \eqref{sols} must now be matched. In this region they are given by
\beq
\psi=B_+ \phi^{\frac 1 2\large(1 + \sqrt{1 -8\alpha} \large)}+B_- \phi^{\frac 1 2 \large(1 - \sqrt{1 -8\alpha} \large)},
\eeq
where $B_+/B_-=\exp(i \zeta)$ with a real phase $\zeta$. The phases extracted from the solutions \eqref{soll} and \eqref{sols} can differ only by the angle $2\pi n$, where $n$ is an integer. The nodeless wave function $\psi$ is found for $n=1$. The matching of the phases $\zeta$ in the intermediate region with $n=1$ gives rise to Eq. \eqref{alpha18}.

\subsection{Analytic calculation near $\alpha=0$}
Here we show how the mass ratio for the minus-branch can be determined analytically close to the non-interacting point $\alpha=0$. In this case identical bosons and fermions must be  treated differently.

In the case of identical bosons the critical mass ratio vanishes at $\alpha=0$. We can thus start from the analytical solution of the angular equation \eqref{polSch} at $\Delta=\pi/4$
\beq \label{wfpi4}
\psi\propto \sin^{1/2-\sqrt{1/4-\alpha}}(\theta+\frac \pi 4)\sin^{1/2-\sqrt{1/4-\alpha}}(\theta-\frac \pi 4)
\eeq
with the positive eigenvalue
\beq\label{eq:Lambda}
-s^2=2(1-2\sqrt{1/4-\alpha}-2\alpha)=4\alpha^2+O(\alpha^3).
\eeq
Let us now take into account a small deviation of $\Delta$ from $\pi/4$ perturbatively. In order to do this we redefine the angle $\theta= 4\Delta \tilde{\theta}/\pi$ such that the configurational space is the fixed interval $\tilde{\theta}\in (0,\pi/4)$ independent of $\Delta$. After this redefinition, Eq.~\eqref{polSch} becomes
\beq\label{eq:RescaledpolSch}
\Big\{-\partial_{\tilde\theta}^2-\frac{\alpha \rho^2}{ \sin^2[\rho(\frac \pi 4+\tilde{\theta})]}-\frac{\alpha \rho^2}{\sin^2[\rho(\frac \pi 4-\tilde{\theta})]}\Big\}\psi=-s^2\rho^2\psi,
\eeq
where $\rho=4\Delta/\pi>1$. The linear shift of $-s^2(\rho)$ with respect to $\rho-1 \ll 1$ is now determined by the first-order perturbation formula
\begin{equation}\label{eq:Perturbation}
\partial (-s^2\rho^2)/\partial \rho=\langle \psi| v_1| \psi\rangle/\langle \psi| \psi \rangle,
\end{equation}
where $v_1$ is the first derivative of the potential in Eq.~(\ref{eq:RescaledpolSch}) with respect to $\rho$ at $\rho=1$ and $\psi$ is given by Eq.~(\ref{wfpi4}). We have explicitly
\beq
\begin{split}
v_1(\theta)=&-\alpha \sec^3(2\theta)  \\
&\times\big[-3\pi+8\cos(2\theta)+\pi \cos(4\theta)+16\theta \sin(2\theta) \big].
\end{split}
\eeq
In fact, the integrals on the right hand side of Eq.~(\ref{eq:Perturbation}) can be calculated for a constant $\psi$, i.e., setting $\alpha=0$ in Eq.~\eqref{wfpi4}. Then, combining the result with Eq.~(\ref{eq:Lambda}) we finally obtain
\begin{equation}
-s^2\approx 4\alpha^2-2\alpha (\rho-1),
\end{equation}
from which Eq.~\eqref{alpha0b} follows immediately.

For identical fermions the mass ratio diverges at $\alpha=0$. Hence one can follow the procedure described in the previous subsection with the wave function \eqref{sols} now satisfying the minus-branch boundary condition $\psi\sim x^{1/2-\sqrt{1/4-\alpha}}$ at $x=0^+$.  As a result, one finds Eq.~\eqref{alpha0f}. 
\subsection{Born-Oppenheimer approximation for minus-branch}
Here we study the minus branch in the regime $M\gg m$, where the Born-Oppenheimer (BO) approximation can be used and the emergence of the Efimov effect can be intuitively understood. Within this approximation the three-body wave function is factorized
\beq \label{BOfac}
\Psi(R,r)=\Phi(R)\psi(r;R),
\eeq
where the function $\psi(r;R)$ corresponds to a bound state of the light particle in the combined potential of two heavy centers located at $R_1=0$ and $R_2=-R$. This wave function satisfies
\beq \label{Schaa}
\underbrace{-\frac{1}{2m}\Big[\partial_r^2+\frac \alpha {r^2}+\frac \alpha {(r+R)^2} \Big]}_{H_{light}}\psi=\epsilon \psi.
\eeq
Note that the distance $R$ is the only characteristic length scale in Eq.~(\ref{Schaa}). Therefore, if a bound state exists at some $R$, it also exists at any $R$, and its energy equals $\epsilon_R=-\sigma/m R^2$, where $\sigma$ is a positive dimensionless number which depends on $\alpha$ and on the choice of the light-heavy boundary condition.

Next, the Born-Oppenheimer approximation assumes that the light particle adiabatically follows displacements of the heavy ones for which $\epsilon_R$ acts as the effective interaction potential. Thus, the Schr\"odinger equation for the relative motion of the heavy particles reads 
\beq \label{Schheavyaa}
\Big[-\frac{1}{M}\partial_R^2+\epsilon_{\text{GS}}(R)  \Big] \Phi=E \Phi. 
\eeq
The energy spectrum $E$ determines the energy spectrum of the three-body system. The spectrum of Eq.~(\ref{Schheavyaa}) is Efimovian, $E_n\sim e^{-\frac{2\pi n}{s}+\theta}$, where $s\approx \sqrt{\sigma M/m}$. For simplicity, we neglected in Eq. \eqref{Schheavyaa} the diagonal correction to the potential $V_{\text{diag}}(R)= \langle \psi| \partial_R^2 \psi\rangle/M$ which is subleading and does not modify the argument above.


Now we demonstrate that Eq. \eqref{Schaa} actually supports  at least one bound state for the minus-branch condition for any $0<\alpha<1/4$.
To this end, we consider the minus branch variational wave-function which equals
\beq
\psi_\kappa\propto r^{1/2-\sqrt{1/4-\alpha}} \exp(-\kappa r)
\eeq
for $r>0$ and vanishes for $r<0$.
We find that for any $0<\alpha<1/4$ and for a sufficiently small variational parameter $\kappa$
\beq \label{braHket}
\langle \psi_\kappa |H_{light}|\psi_\kappa \rangle<0.
\eeq
This can be seen by evaluating analytically the variational energy (\ref{braHket}) and Taylor expanding it around $\kappa=0$. We thus arrive at the conclusion that for the minus-branch boundary condition the BO approximation demonstrates that the energy spectrum of our three-body system is Efimovian for \emph{any} $0<\alpha<1/4$ in the limit $M\gg m$.

\end{document}